\documentclass[10pt,epsfig]{article}
\usepackage{graphicx}
\usepackage{array}
\usepackage{amsmath,amssymb}
\usepackage[latin1]{inputenc}

\begin{document}

%\documentclass[a4paper]{jpconf}
%\usepackage{graphicx}
%\usepackage{cite}
%\usepackage{amsmath}
%\begin{document}

\title{Ternary generalization of Pauli's principle \\ and the $Z_6$-graded algebras}

\author{Richard Kerner}

\date{}

\maketitle

%\ead{richard.kerner@upmc.fr}

{\it Laboratoire de Physique Th\'eorique de la Mati\`ere Condens\'ee (LPTMC),}

{\it Univesit\'e Pierre et Marie Curie - CNRS UMR 7600}

{\it Tour 23-13, 5-\`eme \'etage, Bo\^{i}te Courrier 121, 4 Place Jussieu, 75005 Paris, FRANCE}

\vskip 0.5cm

\begin{abstract}
{\small We show how the discrete symmetries $Z_2$ and $Z_3$
combined with the superposition principle result in the $SL(2, {\bf C})$-symmetry of quantum states.  
 The role of  Pauli's exclusion principle in the derivation of the
$SL(2, {\bf C})$  symmetry is put forward as the source of the macroscopically observed Lorentz symmetry; 
then it is generalized for the case of the $Z_3$ grading
replacing the usual $Z_2$ grading, leading to ternary commutation relations. 
We discuss the cubic and ternary generalizations of Grassmann algebra.
Invariant cubic forms are introduced, and their symmetry group is shown to be the $SL(2,C)$ group 
The wave equation generalizing the Dirac operator to the $Z_3$-graded case is constructed. 
Its diagonalization leads to a sixth-order equation. The solutions  cannot propagate
 because their exponents always contain non-oscillating real damping factor. 
We show how certain cubic products can propagate nevertheless.
The model suggests the origin of the color $SU(3)$ symmetry.}

\end{abstract}

\section{Introduction}

In modern physics, which was created by scientific giants like  Galileo, Kepler,
Newton and Huygens, the description of the world surrounding us is based on three essential
realms, which are {\it Material bodies}, {\it Forces acting between them} and {\it Space and Time.}
Newton's third law:
\begin{equation}
{\bf a} = \frac{1}{m} \; {\bf F}.
\label{Newton3}
\end{equation}
shows the relation between three different realms which are dominant in our description of physical
world: massive bodies ($m$), force fields responsible for interactions between the bodies 
("${\bf F}$") and space-time relations defining the acceleration ("${\bf a}$").
Similar ingredients are found in physics of fundamental interactions: we speak of elementary 
particles and fields evolving in space and time. 

In the formula (\ref{Newton3}) we deliberately have put the acceleration on the left-hand side, and 
the inverse of mass anf the force on the right-hand side in order to separate the directly observable entity  ${\bf a}$)
from the product of two entities whose definition is much less direct and clear.

Also, by putting the acceleration alone on the left-hand side,
we underline the causal relationship between the
phenomena: the force is the cause of acceleration, and not vice versa.
In modern language, the notion of force is generally replaced by that of a field.
The fact that the three ingredients are related by the equation (\ref{Newton3}) may suggest that perhaps 
only two of them are fundamentally independent, the third one being the consequence of the remaining two.

The three aspects of theories of fundamental interactions can be symbolized by three orthogonal axes,
as shown in following figure, which displays also three choices of pairs of independent properties
from which we are supposed to be able to derive the third one. 
\begin{figure}[hbt!]
\center
  \includegraphics[width=5cm, height=4cm]{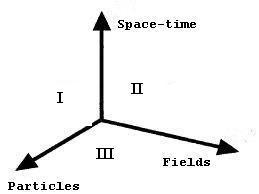}
\label{fig:3realms}  
\caption{The three realms of physical world}
\end{figure}

The attempts to understand physics with only two realms out of three represented  in (\ref{fig:3realms})
have a very long history. They may be divided in three categories, labeled  $I$, $II$
 and $III$ in the Figure above. 

In the category $I$ one can easily recognize Newtonian physics, presenting physical world 
as collection of material bodies (particles) evolving in absolute space and time, interacting at
a distance. Newton considered light being made of tiny particles, too; the notion of fields was
totally absent. Any change in positions and velocities of any massive material object was immediately felt by all
other masses in the entire Universe.

Theories belonging to the category $II$ assume that physical world can be described uniquely as 
a collection of fields evolving in space-time manifold. This approach was advocated by Kelvin, 
Einstein, and later on by Wheeler. As a follower of Maxwell and Faraday, Einstein
believed in the primary role of fields and tried to derive the equations of motion
as characteristic behavior of singularities of fields, or the singularities of the
space-time curvature.

The category $III$ represents an alternative point of view supposing that the existence of matter
is primary with respect to that of the space-time, which becomes an ``emergent" realm - an euphemism
for ``illusion". Such an approach was advocated recently by N. Seiberg and E. Verlinde \cite{Verlinde}. 
 It is true that space-time coordinates cannot be treated on the same
footing as conserved quantities such as energy and momentum; we often forget that they exist
rather as bookkeeping devices, and treating them as real objects is a ``bad habit", as pointed out by D. Mermin
\cite{Mermin}.

Seen under this angle, the idea to derive the geometric 
properties of space-time, and perhaps its very existence, from fundamental symmetries and 
interactions proper to matter's most fundamental building blocks seems quite natural.

 Many of those properties do not require any mention of space and time 
on the quantum mechanical level, as was demonstrated by Born and Heisenberg  
in their version of matrix mechanics, or by von Neumann's formulation of quantum theory in terms
of the $C^*$ algebras \cite{BornJH}, \cite{JvNeumann}.
The non-commutative geometry is another example of formulation
of space-time relationships in purely algebraic terms \cite{MDVRKJM}.

In what follows, we shall choose the latter point of view,
according to which the space-time relations are a consequence of fundamental
{\it discrete symmetries} which characterize the behavior of
matter on the quantum level. In other words, the Lorentz symmetry observed on the macroscopic level,
acting on what we perceive as space-time variables,
is an averaged version of the symmetry group acting in the Hilbert space of quantum states
of fundamental particle systems.

\section{Space-time as emerging realm}

In standard textbooks introducing the Lorentz and Poincar\'e groups the accent is put on
the transformation properties of space and time coordinates, and the invariance of the
Minkowskian metric tensor  $g_{\mu \nu}$. 
But neither its components, nor the space-time coordinates of an observed event can be given an intrinsic
physical meaning; they are not related to any conserved or directly observable quantities.

Under a closer scrutiny, it turns out that only {\it time}  - the proper time of the observer -
can be measured directly. The notion of space variables results from the convenient
description of experiments and observations concerning the propagation of photons, and the
existence of the universal constant $c$.

Consequently, with high enough precision one can infer that the Doppler effect
is relativistic, i.e. the frequency $\omega$ and the wave vector ${\bf k}$
 form an entity that is seen differently by different inertial observers, and passing
 from $\frac{\omega}{c}, {\bf k}$ to  $\frac{\omega'}{c}, {\bf k}'$
is the Lorentz transformation.

Both effects, proving the relativistic formulae
$$ \omega' = \frac{\omega - V k}{\sqrt{1 - \frac{V^2}{c^2}}}, \; \; 
k' = \frac{k - \frac{V}{c^2} \omega}{\sqrt{1 - \frac{V^2}{c^2}}},$$
 have been checked experimentally by Ives and Stilwell in 1937, then confirmed in many more precise experiences.
Reliable experimental confirmations of the validity of Lorentz transformations concern 
measurable quantities such as charges, currents, energies (frequencies)  and momenta (wave vectors) much more 
than the less intrinsic quantities which are the {\it differentials}
 of the space-time variables. 
In principle, the Lorentz transformations could have been established by very precise observations of the 
Doppler effect alone.

It should be stressed that had we only the light at our disposal,
i.e. massless photons propagating with the same velocity $c$, we would infer
that the general symmetry of physical phenomena is the {\it Conformal Group}, and not the Poincar\'e group.
To the observations of light must be added the  {\it the principle of inertia}, i.e. the existence of massive bodies 
moving with speeds lower than $c$, and constant if not sollicited by external influence.

Translated into the modern language of particles and fields
this means that besides the massless photons massive particles must exist, too.
The distinctive feature of such particles is their 
{\it inertial mass}, equivalent with their energy at rest, which can be measured
classically via Newton's law, whose fundamental equation ${\bf a} = \frac{1}{m} \, {\bf F}.$
relates the only  {\it observable quantity}  (using clocks and light rays as measuring rods),
 the acceleration  {\bf a}, with a combination of less evidently defined quantities,  {\it mass} and {\it force},
which is interpreted as a {\it causality relation}, the force being the cause,
and acceleration the effect.

 It turned out soon that the force {\bf F} may symbolize the action 
of quite different physical phenomena like gravitation, electricity or inertia, and is not a primary cause, 
but rather a manner of intermediate bookkeeping.
The more realistic sources of acceleration - or rather of the variation of energy and momenta - are the intensities
of electric, magnetic or gravitational fields.
The differential form of the Lorentz force, combined with the energy conservation of a charged particle
under the influence of electromagnetic field
\begin{equation}
\frac{d {\bf p}}{dt} = q \, {\bf E} + q \, \frac{\bf v}{c} \wedge {\bf B}
\; \; \; \; \; 
\frac{d {\cal{E}}}{dt} = q \, {\bf E} \cdot {\bf v}
\label{EnergyL}
\end{equation}
 is also Lorentz-invariant:
\begin{equation}
d p^{\mu} = \frac{q}{mc} \, F^{\mu}_{\; \; \nu} \, p^{\nu},
\label{Lorentzforce}
\end{equation}
where $p^{\mu} = [p^0, {\bf p} ]$ is the four-momentum and $F^{\mu}_{\; \; \nu} $ is the Maxwell-Faraday
tensor. 

These are the fundamental physical quantities that impose the Lorentz-Poincar\'e group of transformations,
which are imprinted on the {\it dual} space which we call space and time variables. 

\section{Combinatorics and covariance}

Since the advent of quantum theory the discrete view of phenomena observed on microscopic 
level took over the continuum view prevailing in the nineteenth century physics.
The dichotomy between discrete and continuous symmetries
has become a major issue in quantum field theory, of which the fundamental 
 {\it spin and statistics theorem} provides the best illustration.
It stipulates that fields describing particles which obey the Fermi-Dirac statistics, called {\it fermions},
transform under the half-integer representations of the Lorentz group,
whereas fields describing particles which obey the Bose-Einstein statistics,
 {\it bosons}, must transform under the integer representations of the Lorentz group.

The fundamental principle ensuring the existence of electron shells
and the Periodic Table is the {\it exclusion principle} formulated by Pauli: fermionic operators must satisfy the
anti-commutation relations   $\Psi^{a} \Psi^{b} = - \Psi^{b} \Psi^{a}$
which means that two electrons cannot coexist in the same state \cite{Pauli1}.

 With two possible values of spin for the electron in each state, the total number
of states corresponding to each shell (i.e. principal quantum number $n$) becomes $2 \, n^2$.
which is the basis of Mendeleev's periodical system, and of resulting stability of matter \cite{Dyson1}.

Quantum Mechanics started as a non-relativistic theory, but very
soon its relativistic generalization was created.
 As a result, the wave functions in the Schroedinger picture were
required to belong to one of the linear representations of the Lorentz group, which means
that they must satisfy the following {\it covariance principle}:
$$ {\tilde{\psi}} (\tilde{x}) = {\tilde{\psi}} (\Lambda (x)) = S(\Lambda) \, \psi (x).$$

 The nature of the representation  $S (\Lambda)$ determines
the character of the field considered: spinorial, vectorial, tensorial...
As in many other fundamental relations, the seemingly simple equation
$$ {\tilde{\psi}} ({\tilde{x}}) = {\tilde{\psi}} (\Lambda (x)) = S(\Lambda) \, \psi (x).$$
creates a bridge between two totally different realms:  the space-time
 accessible via classical macroscopic observations, and the  Hilbert space of quantum states. 
It can be interpreted in two opposite ways, depending on which
side we consider as the cause, and which one as the consequence.

A question can be asked, what is the cause, and what is the effect,
not only in mathematical terms, but also in a deeper physical sense.
In other words, is the macroscopically observed Lorentz symmetry imposed
on the micro-world of quantum physics,
or maybe it is already present as symmetry of quantum states, and then implemented
 and extended to the macroscopic world in classical limit ? In such a case, the covariance principle
should be written as follows:
$$ \Lambda^{\mu'}_{\mu} (S) \; j^{\mu} = j^{\mu'} (\psi') = j^{\mu'} (S(\psi)),$$ 
In the above formula $ j^{\mu} = \bar{\psi} \gamma^{\mu} \psi $
is the Dirac current, $\psi$ is the electron wave function.

In view of the analysis of the causal chain, it seems more appropriate to write the same transformations
with $\Lambda$ depending on $S$:
\begin{equation}
\psi' (x^{\mu'}) = \psi' (\Lambda^{\mu'}_{\nu} (S) x^{\nu}) = S \psi (x^{\nu})
\label{lambdaes}
\end{equation}
This form of the same relation suggests that the transition 
from one quantum state to another, represented by the  transformation $S$ 
is the primary cause that implies the transformation
of observed quantities such as the electric $4$-current, and as a final consequence, the apparent
transformations of time and space intervals measured with classical physical devices.

The Pauli exclusion principle gives a hint about how it might work.
In its simplest version, it introduces an anti-symmetric form on the Hilbert space describing 
electron's states:
$$\epsilon^{\alpha \beta} = - \epsilon^{\beta \alpha}, \; \;  \alpha, \beta = 1,2; \; \; \; \epsilon^{12} =1,$$
Now, if we require that Pauli's principle must apply 
independently of the choice of a basis
in Hilbert space, i.e. that after a linear transformation we get
$$\epsilon^{\alpha' \beta'} = S^{\alpha'}_{\alpha} S^{\beta' }_{\beta} \epsilon^{\alpha \beta}
= - \epsilon^{\beta' \alpha'}, 
\; \; \; \epsilon^{1' 2'} = 1,$$
then the matrix $S^{\alpha' }_{\alpha}$ must have the determinant equal to $1$,
which defines the $SL(2, {\bf C})$ group.

The existence of two internal degrees of freedom had to be taken
into account in fundamental equation defining the relationship between basic operators
acting on electron states. 
To acknowledge this, Pauli proposed the simplest equation expressing the relation between 
the energy, momentum and spin:
\begin{equation}
E \, \psi = mc^2 \psi + {\boldsymbol \sigma} \cdot {\bf p} \, \psi.
\label{Paulieq1}
\end{equation}

 The existence of  {\it anti-particles} (in this case the positron),
suggests the use of the non-equivalent representation of  $SL(2, {\bf C})$ group by means
of complex conjugate matrices. along with the time reversal, the Dirac equation can be now constructed. It is
invariant under the Lorentz group.
\begin{equation}
E \psi_{+} = mc^2 \psi_{+} + {\boldsymbol \sigma} \cdot {\bf p} \psi_{-}, \; \; \; \; 
- E \psi_{-} = mc^2 \psi_{-} + {\boldsymbol \sigma} \cdot {\bf p} \psi_{+}
%$$i \hbar \frac{\partial }{\partial t} \psi_{+} - mc^2 \psi_{+} = 
% i \hbar {\bf \sigma} \cdot {\bf \nabla} \psi_{-},$$
%\begin{equation}
%- i \hbar \frac{\partial }{\partial t} \psi_{-} - mc^2 \psi_{-} = 
%- i \hbar {\bf \sigma} \cdot {\bf \nabla} \psi_{+},
\label{Diraccoupled1}
\end{equation}
Although mathematically the two formulations are equivalent, it seems more plausible 
that the Lorentz group resulting from the averaging of the action of the 
$SL(2, {\bf C})$ in the Hilbert space of states
contains less information than the original double-valued representation which is a consequence of the
particle-anti-particle symmetry, than the other way round.
In what follows, we shall draw physical consequences from this approach, concerning the strong interactions
in the first place.
 
In purely algebraical terms Pauli's exclusion principle amounts to the anti-symmetry of 
wave functions describing two coexisting particle states.
The easiest way to see how the principle
works is to apply Dirac's formalism in which wave functions of particles in given state are
obtained as products between the ``bra" and ``ket" vectors.
 Consider the wave function of a particle in the state $\mid x >$, 
\begin{equation}
\Phi (x) = < \psi \mid x>.
\label{x-state}
\end{equation}
 A two-particle state of  $(\mid x>, \mid y )$ is a tensor product
\begin{equation}
\mid \psi > = \sum \, \Phi (x,y) \, ( \mid x> \otimes \mid y>).
\label{xy-state}
\end{equation}
If the wave function $\Phi (x,y)$ is anti-symmetric, i.e. if it satisfies
\begin{equation}
\Phi (x,y) = - \Phi (y,x),
\label{antisymPhi}
\end{equation}
then  $\Phi (x,x) = 0$ and such states have vanishing probability.

Conversely, suppose that $\Phi (x,x)$ does
 vanish. This remains valid in any basis  provided
the new basis  $\mid x'>, \; \mid y'>$ was obtained from the former one via unitary transformation.

 Let us form an arbitrary state being a linear combination of $\mid x>$ and $\mid y>$,
$$ \mid z> = \alpha \mid x > + \beta \mid y >, \; \; \; \alpha, \beta \in {\bf C},$$
 and let us form the wave function of a tensor product of such a state with itself:
\begin{equation}
\Phi (z,z) = < \psi \mid ( \alpha \mid x > + \beta \mid y > ) \otimes ( \alpha \mid x > + \beta \mid y > ),
\label{Atensor}
\end{equation}
which develops as follows:
$$ \alpha^2 \, < \psi \mid x,x> + \alpha \beta  < \psi \mid x,y > $$
$$+ \beta \alpha \, < \psi \mid y,x > + \beta^2 \,  < \psi \mid y,y > =  $$
\begin{equation}
 = \alpha^2 \,  \Phi (x,x) + \alpha \beta \,  \Phi (x,y) 
+ \beta \alpha \, \Phi (y,x) + \beta^2 \, \Phi (y,y).
\label{Adeveloped}
\end{equation}
Now, as  $ \Phi (x,x) = 0$ and $\Phi (y,y)=0$, the sum of remaining 
two terms will vanish if and only if 
(\ref{antisymPhi}) is
 satisfied, i.e. if $\Phi (x,y)$  is anti-symmetric in its two arguments.

 After second quantization, when the states are obtained with creation 
and annihilation operators acting
on the vacuum, the anti-symmetry is encoded in the anti-commutation relations
%\begin{equation}
%a^{\dagger} (x) a^{\dagger} (y) + a^{\dagger} (y) a^{\dagger} (x) = 0.
%\label{anticommutation}
%\end{equation}
\begin{equation}
\psi (x) \psi(y) + \psi (y) \psi (x) = 0 \; \; \; 
\label{anticompsi}
\end{equation}
where $\psi(x) \mid 0> = \mid x >$.

 According to present knowledge, the ultimate undivisible and undestructible
constituents of matter, called {\it atoms} by ancient Greeks, are in fact
the  QUARKS, carrying fractional electric charges and baryonic numbers, 
two features that appear to be  undestructible and conserved under any circumstances.

Taking into account that quarks evolve inside nucleons as almost point-like
objects, one may wonder how the notions of space and time still apply in these conditions ?
Perhaps in this case, too, the Lorentz invariance can be derived from some more fundamental
 {\it discrete} symmetries underlying the interactions between quarks ?
 If this is the case, then the symmetry $Z_3$ must play a fundamental role.

In Quantum Chromodynamics quarks are considered as fermions, endowed  with spin $\frac{1}{2}$.
Only {\it three}
quarks or anti-quarks can coexist inside a fermionic baryon (respectively, anti-baryon), and a pair 
quark-antiquark can form a meson with integer spin.
Besides, they must belong to different
{\it colors}, also a three-valued set. There are two quarks in the first generation, 
$u$ and $d$ (``up" and ``down"), which may be considered as two states of a more general object,
just like proton and neutron in $SU(2)$ symmetry are two isospin components of a nucleon doublet.
%While there is place for {\it two}
%quarks in the same $u$-state or $d$-state, but not three.

This suggests that a convenient generalization of Pauli's exclusion principle 
would be that no {\it three} quarks in the same state can be present in a nucleon.

Let us require then the vanishing of wave functions representing the tensor product of {\it three}
(but not necessarily two) identical states. That is, we require that 
$\Phi (x,x,x) = 0$ for any state  $\mid x>$. 
As in the former case, consider an arbitrary superposition of 
three different states, $\mid x>, \; \mid y>$ and $\mid z>$,
$$ \mid w> = \alpha \mid x> + \beta \mid y> + \gamma \mid z>$$
and apply the same criterion, $\Phi (w,w,w) = 0$. 

We get then, after developing the tensor products,
$$ \Phi (w,w,w) = \alpha^3 \Phi(x,x,x) + \beta^3 \Phi( y,y,y) + \gamma^3 \Phi (z,z,z) $$
$$+ \alpha^2 \beta [ \Phi(x,x,y) + \Phi(x,y,x) + \Phi(y,x,x)] +
\gamma \alpha^2  [\Phi(x,x,z) + \Phi(x,z,x) + \Phi(z,x,x)] $$
$$+ \alpha \beta^2  [\Phi(y,y,x) + \Phi(y,x,y) + \Phi(x,y,y)] 
+ \beta^2 \gamma [\Phi(y,y,z) + \Phi(y,z,y) + \Phi(z,y,y)]$$
$$+ \beta \gamma^2 [\Phi(y,z,z) + \Phi(z,z,y) + \Phi(z,y,z)] 
+ \gamma^2 \alpha [\Phi(z,z,x) + \Phi(z,x,z) + \Phi(x,z,z)] $$
$$+ \alpha \beta \gamma [ \Phi(x,y,z) + \Phi(y,z,x) + \Phi(z,x,y) + \Phi(z,y,x)
+ \Phi(y,x,z) + \Phi(x,z,y) ]=0.$$ 

%$$ < \psi \mid ( \alpha \mid x > + \beta \mid y > + \gamma \mid z> ) \otimes 
%( \alpha \mid x> + \beta \mid y + \gamma \mid z> ) \otimes ( \alpha \mid x > + \beta \mid y >
%+ \gamma \mid z> )=$$

The terms $\Phi (x,x,x), \; \Phi (y,y,y)$ and $\Phi (z,z,z)$ do vanish by 
virtue of the original assumption; in what remains, combinations preceded by various
powers of independent numerical coefficients $\alpha, \beta $
 and $\gamma$, must vanish separately. 

This is achieved if the following $Z_3$ symmetry is imposed on our wave functions:
$$\Phi(x,y,z) = j \, \Phi (y,z, x) = j^2 \, \Phi (z,x,y).$$
with  $j=e^{\frac{2 \pi i}{3}}, \; \; j^3 = 1, \; \; j+j^2 +1 =0.$

Note that the complex conjugates of functions $\Phi (x,y,z)$ transform under cyclic permutations
of their arguments with $j^2 = \bar{j}$ replacing $j$ in the above formula 
$$\Psi(x,y,z) = j^2 \, \Psi (y,z,x) = j \, \Psi (z,x,y).$$
Inside a hadron, not two, but three quarks in different states (colors) can coexist. 

After second quantization, when the fields become operator-valued,
an alternative {\it cubic commutation relations} seems to
be more appropriate:

Instead of $ \Psi^a \Psi^b = (-1) \, \Psi^b \Psi^a$
we can introduce $ \theta^A \theta^B \theta^C = j \, \theta^B \theta^C \theta^A = 
j^2 \,  \theta^C \theta^A \theta^B ,$ with $j = e^{\frac{2 \pi i}{3}}$

\section{Quark algebra}

Our aim now is to derive the space-time symmetries from minimal assumptions concerning the
properties of the most elementary constituents of matter, and the best candidates for
these are quarks.

To do so, we should explore algebraic structures that would privilege {\it cubic} or
{\it ternary} relations, in other words, find 
appropriate  cubic or  ternary algebras reflecting the most
important properties of quark states.
The minimal requirements for the definition of quarks at the initial stage of model building
are the following:

{\it i} ) 
The mathematical entities representing the quarks form a linear space over complex numbers, 
so that we could form their linear combinations with complex coefficients.

{\it ii} ) 
They should also form an associative algebra, so that we could form their multilinear 
combinations;

{\it iii} ) 
There should exist two isomorphic algebras of this type corresponding to quarks
and anti-quarks, and the conjugation that maps one of these algebras onto
another, ${\cal{A}} \rightarrow {\bar{\cal{A}}}$.

{\it iv} ) 
The three quark (or three anti-quark) and the quark-anti-quark combinations 
should be distinguished in a certain way, for example, they should form a subalgebra
in the enveloping algebra spanned by the generators. 

The fact that hadrons obeying the Fermi statistics (protons and neutrons, to begin with) are composed
of three quarks raises naturally the question how their quantum states respond to permutations
between these elementary components. 

 The symmetric group  $S_3$ containing all permutations of three different elements
is a special case among all symmetry groups $S_N$. It is  
the first in the row to be non-abelian, and the last one that possesses a faithful representation
in the complex plane  ${\bf C}^1$.
It contains six elements, and can be generated with only two elements, corresponding to one cyclic
and one odd permutation, e.g. $(abc) \rightarrow (bca)$, and $(abc) \rightarrow (cba)$.
All permutations can be represented as different operations on complex numbers as follows.

 Let us denote the primitive third root of unity by  $j = e^{2 \pi i/3}$.

 The cyclic abelian subgroup $Z_3$ contains three elements corresponding to the three
cyclic permutations, which can be represented via multiplication by  $j$, $j^2$ and  $j^3 =1$ (the identity).
\begin{equation}
\begin{pmatrix}
ABC \cr ABC
\end{pmatrix}
\rightarrow {\bf 1}, \, \ \ \, 
\begin{pmatrix}
ABC \cr BCA
\end{pmatrix}
\rightarrow {\bf j}, \, \ \ \,
\begin{pmatrix}
ABC \cr CAB
\end{pmatrix}
\rightarrow {\bf j^2},
\label{permutationseven}
\end{equation}

Odd permutations must be represented by idempotents, i.e. by operations whose square
is the identity operation. We can make the following choice:
{\small
\begin{equation}
\begin{pmatrix}
ABC \cr CBA
\end{pmatrix}
\rightarrow ({\bf z \rightarrow {\bar{z}}}), \, \ \ \, 
\begin{pmatrix}
ABC \cr BAC
\end{pmatrix}
\rightarrow ({\bf z \rightarrow {\hat{z}}}), \, \ \ \,
\begin{pmatrix}
ABC \cr CBA
\end{pmatrix}
\rightarrow ({\bf z \rightarrow z^{*}}),
\label{permutationsodd}
\end{equation}
Here the bar $({\bf z \rightarrow {\bar{z}}})$ denotes the complex conjugation, i.e. the reflection
in the real line, the hat ${\bf z \rightarrow {\hat{z}}}$ denotes the reflection in the root  $j^2$,
and the star ${\bf z \rightarrow z^{*}}$ the reflection in the root  $j$.
The six operations close in a non-abelian group with six elements. However, if it acts on three objects 
out of which two are identical, e.g. $(AAB)$, then odd permutations give the same result as even ones,
so that only the $Z_3$ cyclic abelian group is operating, 
With this in mind, let us define the following $Z_3$-graded algebra introducing $N$ generators 
spanning a linear space over complex numbers, satisfying the following cubic relations:

\begin{equation}
\theta^A \theta^B \theta^C = j \, \theta^B \theta^C \theta^A = j^2 \, \theta^C \theta^A \theta^B,
\label{ternary1}
\end{equation}
with $j = e^{2 i \pi/3}$, the primitive root of $1$. We have obviously 
$1+j+j^2 = 0$} \; \; and ${\bar{j}} = j^2$.

We shall also introduce a similar set of {\it conjugate} generators, ${\bar{\theta}}^{\dot{A}}$,
$\dot{A}, \dot{B},... = 1,2,...,N$, satisfying similar condition with $j^2$ replacing $j$:
\begin{equation}
{\bar{\theta}}^{\dot{A}} {\bar{\theta}}^{\dot{B}} {\bar{\theta}}^{\dot{C}} = 
j^2 \, {\bar{\theta}}^{\dot{B}} {\bar{\theta}}^{\dot{C}} {\bar{\theta}}^{\dot{A}} 
= j \, {\bar{\theta}}^{\dot{C}} {\bar{\theta}}^{\dot{A}} {\bar{\theta}}^{\dot{B}},
\label{ternary2}
\end{equation}
Let us denote this algebra by ${\bf{\cal{A}}}$.

We shall endow it with a natural $Z_3$ grading, considering the generators $\theta^A$
as grade $1$ elements, their conjugates  ${\bar{\theta}}^{\dot{A}}$ being of grade $2$.
 The grades add up modulo $3$; the products $\theta^{A} \theta^{B}$ span a linear
subspace of grade $2$, and the cubic products $\theta^A \theta^B \theta^C$ are of grade $0$.

 Similarly, all quadratic expressions in conjugate generators, ${\bar{\theta}}^{\dot{A}} {\bar{\theta}}^{\dot{B}}$
are of  grade $2 + 2 = 4_{mod \, 3} = 1$, whereas their cubic products are again of grade $0$, like
the cubic products od $\theta^A$'s.

Combined with the associativity, these cubic relations impose finite dimension on the
algebra generated by the $Z_3$ graded generators. As a matter of fact, cubic expressions are the
highest order that does not vanish identically. The proof is immediate:
\begin{equation}
\theta^A \theta^B \theta^C \theta^D = j \, \theta^B \theta^C \theta^A \theta^D =
j^2 \, \theta^B \theta^A \theta^D \theta^C 
= j^3 \, \theta^A \theta^D \theta^B \theta^C =
j^4 \, \theta^A \theta^B \theta^C \theta^D,
\label{quartic1}
\end{equation}
and because $j^4 = j \neq 1$, the only solution is $\theta^A \theta^B \theta^C \theta^D = 0.$

The total dimension of the algebra defined via the cubic relations (\ref{ternary1})
is equal to $N + N^2 + (N^3 - N)/3$: the $N$ generators of grade $1$, the $N^2$ independent
products of two generators, and $(N^3-N)/3$ independent cubic expressions, because the cube
of any generator must be zero by virtue of (\ref{ternary1}), and the remaining $N^3-N$ ternary
products are divided by $3$, also by virtue of the constitutive relations (\ref{ternary1}).

The conjugate generators ${\bar{\theta}}^{\dot{B}}$ span an algebra ${\bf{\bar{\cal{A}}}}$ isomorphic 
with ${\bf{\cal{A}}}$.

If we want the products between the generators $\theta^A$ and the conjugate ones ${\bar{\theta}}^{\dot{B}}$
to be included into the greater algebra spanned by both types of generators, we should
consider all possible products, between both types of generators, which will span
the resulting algebra  ${\cal{A}} \otimes {\bar{\cal{A}}}$.

The fact that the conjugate generators are endowed with grade $2$ could suggest that
they behave just like the products of two ordinary generators $\theta^A \theta^B$.
However, such a choice does not enable one to make a clear distinction between the
conjugate generators and the products of two ordinary ones, and it would be much better,
to be able to make the difference. 

Due to the binary nature of the products, another
choice is possible, namely, to require the following commutation relations:
\begin{equation}
\theta^{A} {\bar{\theta}}^{\dot{B}} = - j \, {\bar{\theta}}^{\dot{B}} \theta^{A}, \, \ \ \, \ \ 
{\bar{\theta}}^{\dot{B}} \theta^{A} = - j^2 \,\theta^{A} {\bar{\theta}}^{\dot{B}},
\label{commutation2}
\end{equation}
In fact, introducing the ``minus" sign, i.e. the multiplication by $-1$, we extend the discrete symmetry group
acting on our algebra to the product $Z_3 \times Z_2$. It is easy to prove that this product is isomorphic
with the cyclic group $Z_6$. The choice of commutation relations (\ref{commutation2}) leads to the anticommutation property between the conjugate
cubic monomials:
\begin{equation}
\left ( \theta^A \theta^B \theta^C \right) \left({\bar{\theta}}^{\dot{D}} {\bar{\theta}}^{\dot{E}} {\bar{\theta}}^{\dot{F}} \right)
= - \left({\bar{\theta}}^{\dot{D}} {\bar{\theta}}^{\dot{E}} {\bar{\theta}}^{\dot{F}} \right) \left ( \theta^A \theta^B \theta^C \right) ,
\label{anticub}
\end{equation}
characteristic for the fermions. This is another hint towards the possibility of forming anti-commuting fermionic variables 
with cubic combinations of our ``quark" operators. 

\section{Two-generator algebra and its invariance group}

The three quarks constituting hadrons (the latter behaving as fermions) are found in two states, ``up" and ``down", designed by
$u$ and $d$, endowed with fractional electric charges, $+\frac{2}{3}$ for the $u$-quark and $- \frac{1}{3}$ for the $d$-quark.
Therefore the product state $uud$ will represent a proton (electric charge $+1$), whilst the combination $udd$ having zeo electric
charge represents a neutron. We shall therefore reduce the number of generators of our $Z_3$-graded algenra representing quark operators,
to the minimal number, i.e. two generators only.

Let us consider the simplest case of cubic algebra with two generators, $A, B,...= 1,2$.
Its grade $1$ component contains just these two elements, $\theta^1$ and $\theta^2$;
its grade $2$ component contains four independent products, 
$\theta^1 \theta^1, \, \theta^1 \theta^2, \, \theta^2 \theta^1, \, \ \  {\rm and} \, \ \  \theta^2 \theta^2. $
Finally, its grade $0$ component (which is a subalgebra) contains the unit element $1$ and the two linearly
independent cubic products, 
$\theta^1 \theta^2 \theta^1 = j \, \theta^2 \theta^1 \theta^1 = j^2 \, \theta^1 \theta^1 \theta^2 \; \; \; 
{\rm and} \; \; \;
\theta^2 \theta^1 \theta^2 = j \, \theta^1 \theta^2 \theta^2 = j^2 \, \theta^2 \theta^2 \theta^1.$
with similar two independent combinations of conjugate generators ${\bar{\theta}}^{\dot{A}}.$

Let us consider multilinear forms defined on the algebra ${\cal{A}} \otimes {\bar{\cal{A}}}$.
Because only cubic relations are imposed on products in ${\cal{A}}$ and in $\bar{\cal{A}}$,
and the binary relations on the products of ordinary and conjugate elements, we shall fix
our attention on tri-linear and bi-linear forms, conceived as mappings of ${\cal{A}} \otimes {\bar{\cal{A}}}$
into certain linear spaces over complex numbers.
Consider a tri-linear form $\rho^{\alpha}_{ABC}$. We shall call this form $Z_3$-invariant if 
we can write, by virtue of (\ref{ternary1}):
$$
\rho^{\alpha}_{ABC} \, \theta^A  \theta^B  \theta^C = \frac{1}{3} \, \biggl[ \rho^{\alpha}_{ABC} \, 
\theta^A  \theta^B  \theta^C
+ \rho^{\alpha}_{BCA} \, \theta^B  \theta^C  \theta^A +  \rho^{\alpha}_{CAB} \, \theta^C  \theta^A  \theta^B \biggr] =$$
$$
= \frac{1}{3} \, \biggl[ \rho^{\alpha}_{ABC} \, \theta^A  \theta^B  \theta^C
+ \rho^{\alpha}_{BCA} \, (j^2 \, \theta^A  \theta^B  \theta^C) +  \rho^{\alpha}_{CAB} \, 
j \, (\theta^A  \theta^B  \theta^C) \biggr],$$
From this it follows that we should have
\begin{equation}
\rho^{\alpha}_{ABC} \, \, \theta^A  \theta^B  \theta^C  = \frac{1}{3} \, \biggl[ \rho^{\alpha}_{ABC}
+ j^2 \, \rho^{\alpha}_{BCA} + j \,  \rho^{\alpha}_{CAB} \biggr] \, \theta^A  \theta^B  \theta^C ,
\label{defrhomatrix2}
\end{equation}
from which we get the following properties of the $\rho$-cubic matrices:
\begin{equation}
\rho^{\alpha}_{ABC} = j^2 \, \rho^{\alpha}_{BCA} = j \,  \rho^{\alpha}_{CAB}.
\label{defrhomatrix3}
\end{equation}
Even in this minimal and discrete case, there are covariant and contravariant
indices: the lower and the upper indices display the inverse transformation property. If a given
cyclic permutation is represented by a multiplication by $j$ for the upper indices, the same permutation performed 
on the lower indices is represented by multiplication by the inverse, i.e. $j^2$, so that they 
compensate each other.

Similar reasoning leads to the definition of the conjugate forms 
$ {\bar{\rho}}^{{\dot{\alpha}}}_{{\dot{C}}{\dot{B}}{\dot{A}}}$
satisfying the relations similar to (\ref{defrhomatrix3}) with $j$ replaced be its conjugate, $j^2$:
\begin{equation}
{\bar{\rho}}^{{\dot{\alpha}}}_{{\dot{A}}{\dot{B}}{\dot{C}}} = j \, 
{\bar{\rho}}^{{\dot{\alpha}}}_{{\dot{B}}{\dot{C}}{\dot{A}}}
= j^2 \, {\bar{\rho}}^{{\dot{\alpha}}}_{{\dot{C}}{\dot{A}}{\dot{B}}} 
\label{defrhomatrix4}
\end{equation}
In the simplest case of two generators, the $j$-skew-invariant forms have
only two independent components:
$$\rho^{1}_{121} = j \, \rho^{1}_{211}
= j^2 \, \rho^{1}_{112}, \; \; \; \; \rho^{2}_{212} = j \, \rho^{2}_{122}
= j^2 \, \rho^{2}_{221},$$
and we can set
$$\rho^{1}_{121} = 1, \, \, \rho^{1}_{211} = j^2,  \, \, \rho^{1}_{112} = j, \; \; \; \; 
\rho^{2}_{212} = 1, \,  \, \rho^{2}_{122} = j^2, \,  \, \rho^{2}_{221} = j.$$
The constitutive cubic relations between the generators of the $Z_3$ graded algebra
can be considered as intrinsic if they are conserved after linear transformations with commuting 
(pure number) coefficients, i.e. if they are independent of the choice of the basis.

Let $U^{A'}_A$ denote a non-singular $N \times N$ matrix, transforming the generators
$\theta^A$ into another set of generators, $\theta^{B'} = U^{B'}_B \, \theta^B$.

We are looking for the solution of the covariance condition for the $\rho$-matrices:
\begin{equation}
S^{{\alpha}'}_{\beta} \, \rho^{{\beta}}_{ABC} = U^{A'}_{A} \, U^{B'}_B \, U^{C'}_C \, 
\rho^{{\alpha}'}_{A' B' C'}.
\label{covtrans1}
\end{equation}
Now, $\rho^{1}_{121} = 1$, and we have two equations corresponding to the choice of values of the index $\alpha'$
equal to $1$ or $2$. For $\alpha' = 1'$ the $\rho$-matrix on the right-hand side is $\rho^{1'}_{A' B' C'}$,
which has only three components,
$$\rho^{1'}_{1' 2' 1'}=1, \, \ \ \, \rho^{1'}_{2' 1' 1'}=j^2, \, \ \ \, \rho^{1'}_{1' 1' 2'}=j, $$
which leads to the following equation:
{\small
\begin{equation*}
S^{1'}_{1} = U^{1'}_{1} \, U^{2'}_2 \, U^{1'}_1 + j^2 \, U^{2'}_{1} \, U^{1'}_2 \, U^{1'}_1 
+ j \, U^{1'}_{1} \, U^{1'}_2 \, U^{2'}_1 = U^{1'}_{1} \, (U^{2'}_2 \, U^{1'}_1 - U^{2'}_{1} \, U^{1'}_2),
\label{invariant1}
\end{equation*}} 
because $j^2 + j = - 1$.

For the alternative choice $\alpha' = 2'$ the $\rho$-matrix on the right-hand side is $\rho^{2'}_{A' B' C'}$,
whose three non-vanishing components are
$$\rho^{2'}_{2' 1' 2'}=1, \, \ \ \, \rho^{2'}_{1' 2' 2'}=j^2, \, \ \ \, \rho^{2'}_{2' 2' 1'}=j. $$
The corresponding equation becomes now:
{\small
\begin{equation*}
S^{2'}_{1} = U^{2'}_{1} \, U^{1'}_2 \, U^{2'}_1 + j^2 \, U^{1'}_{1} \, U^{2'}_2 \, U^{2'}_1 
+ j \, U^{2'}_{1} \, U^{2'}_2 \, U^{1'}_1 = U^{2'}_{1} \, (U^{1'}_2 \, U^{2'}_1 - U^{1'}_{1} \, U^{2'}_2),
\label{invariant2}
\end{equation*} }
The remaining two equations are obtained in a similar manner. We choose now the three lower indices
on the left-hand side equal to another independent combination, $(212)$. Then the $\rho$-matrix on the
left hand side must be $\rho^2$ whose component $\rho^2_{212}$ is equal to $1$. This leads to the
following equation when $\alpha' = 1'$:
{\small
\begin{equation*}
S^{1'}_{2} = U^{1'}_{2} \, U^{2'}_1 \, U^{1'}_2 + j^2 \, U^{2'}_{2} \, U^{1'}_1 \, U^{1'}_2
+ j \, U^{1'}_{2} \, U^{1'}_1 \, U^{2'}_2 = U^{1'}_{2} \, (U^{1'}_2 \, U^{2'}_1 - U^{1'}_{1} \, U^{2'}_2),
\label{invariant3}
\end{equation*}}
and the fourth equation corresponding to $\alpha' = 2'$ is:
{\small \begin{equation*}
S^{2'}_{2} = U^{2'}_{2} \, U^{1'}_1 \, U^{2'}_2 + j^2 \, U^{1'}_{2} \, U^{2'}_1 \, U^{2'}_2
+ j \, U^{2'}_{2} \, U^{2'}_1 \, U^{1'}_2 = U^{2'}_{2} \, (U^{1'}_1 \, U^{2'}_2 - U^{2'}_{1} \, U^{1'}_2).
\label{invariant4}
\end{equation*}}
The determinant of the $2 \times 2$ complex matrix $U^{A'}_B$ appears everywhere 
on the right-hand side.
\begin{equation}
S^{2'}_{1} = - U^{2'}_{1} \, [det(U)],
\label{Lambda21}
\end{equation} 
The remaining two equations are obtained in a similar manner, resulting in the following:
\begin{equation}
S^{1'}_{2} = - U^{1'}_{2} \, [det(U)], \, \ \ \, \ \ S^{2'}_{2} = U^{2'}_{2} \, [det(U)].
\label{invariant34}
\end{equation}
The determinant of the $2 \times 2$ complex matrix $U^{A'}_B$ appears everywhere 
on the right-hand side.
Taking the determinant of the matrix $S^{{\alpha}'}_{\beta}$ one gets immediately
\begin{equation}
{\rm det} \, ( S ) = [ {\rm det} \, (U) ]^3.
\label{detLambdaU}
\end{equation}
However, the $U$-matrices on the right-hand side are defined 
only up to the phase, which due to the
cubic character of the covariance relations (\ref{invariant1} - \ref{invariant34}), 
and they can take on three different values: $1$, $j$ or $j^2$,
i.e. the matrices $j \, U^{A'}_B$ or $j^2 \,  U^{A'}_B$ satisfy the same relations
as the matrices $U^{A'}_B$ defined above. 
The determinant of $U$ can take on the values $1, \, j \,$ or $j^2$ if  $det (S) = 1$
But for the time being, we have no reason yet to impose the unitarity condition. It can be
derived from the conditions imposed on the invariance and duality.of binary relations between $\theta^A$ and
their conjugates ${\bar{\theta}}^{\dot{B}}$.

In the Hilbert space of spinors the $SL(2, {\bf C})$ action conserved naturally two anti-symmetric tensors, 
$$\varepsilon_{\alpha \beta} \; \; \; {\rm and} \; \; \; \varepsilon_{\dot{\alpha} \dot{\beta}} \; \; {\rm and \; their \; duals} \; \; 
\varepsilon^{\alpha \beta} \; \; \; {\rm and} \; \; \; \varepsilon^{\dot{\alpha} \dot{\beta}}.$$
Spinorial indeces thus can be raised or lowered using these fundamental
$SL(2, {\bf C})$ tensors:
$$\psi_{\beta} = \epsilon_{\alpha \beta} \, \psi^{\alpha}, \; \; \; 
\psi^{\dot{\delta}} = \varepsilon^{\dot{\delta} \dot{\beta}} \; \psi_{\dot{\beta}}.$$

In the space of quark states similar invariant form can be introduced, too. Theere is
only one alternative: either the Kronecker delta, or the anti-symmetric $2$-form $\varepsilon$.
Supposing that our cubic combinations of quark states behave like fermions, 
there is no choice left: if we want to define the duals of cubic forms $\rho^{\alpha}_{ABC}$ 
displaying the same symmetry properties, we must impose the covariance principle as follows:
$$\epsilon_{\alpha \beta} \; \rho^{\alpha}_{ABC} = \varepsilon_{AD} \varepsilon_{BE} \varepsilon_{CG}
\; \rho_{\beta}^{DEG}.$$
The requirement of the invariance of tensor $\varepsilon_{AB}$, $A,B = 1,2$
 with respect to the change of basis
of quark states leads to the condition ${\rm det} U = 1$, i.e. again to the $SL(2, {\bf C})$ group. 

A similar covariance requirement can be formulated with respect to the set of $2$-forms
mapping the quadratic quark-anti-quark combinations into a four-dimensional linear real space.
As we already saw, the symmetry (\ref{commutation2}) imposed on these expressions reduces their
number to four. Let us define two quadratic forms, $\pi^{\mu}_{A {\dot{B}}}$
 and its conjugate  ${\bar{\pi}}^{\mu}_{{\dot{B}} A}$ 
\begin{equation}
\pi^{\mu}_{A {\dot{B}}} \, \theta^A {\bar{\theta}}^{\dot{B}} 
\; \; \; {\rm and} \; \; \;  {\bar{\pi}}^{\mu}_{{\dot{B}} A}
\, {\bar{\theta}}^{\dot{B}} \theta^A.
\label{pisymmetry1}
\end{equation}
The Greek indices $\mu, \nu...$ take on four values, and we shall label them $0,1,2,3$.
The four tensors $\pi^{\mu}_{A {\dot{B}}}$ and their hermitina conjugates 
${\bar{\pi}}^{\mu}_{{\dot{B}} A}$ define a bi-linear mapping from the product of
quark and anti-quark cubic algebras into a linear four-dimensional vector space, whose
structure is not yet defined.
Let us impose the following invariance condition:
\begin{equation} 
\pi^{\mu}_{A {\dot{B}}} \, \theta^A {\bar{\theta}}^{\dot{B}} = {\bar{\pi}}^{\mu}_{{\dot{B}} A}
{\bar{\theta}}^{\dot{B}} \theta^A.
\label{invbin}
\end{equation}
It follows immediately from (\ref{commutation2}) that
\begin{equation}
\pi^{\mu}_{A {\dot{B}}} = - j^2 \, {\bar{\pi}}^{\mu}_{{\dot{B}} A}.
\label{pisymmetry2}
\end{equation}
Such matrices are non-hermitian, and they can be realized by the following substitution:
\begin{equation}
\pi^{\mu}_{A {\dot{B}}} = j^2 \, i \, {\sigma}^{\mu}_{A {\dot{B}}}, \, \ \ \,
{\bar{\pi}}^{\mu}_{{\dot{B}} A} = - j \, i \, {\sigma}^{\mu}_{{\dot{B}} A}
\label{pidefinition2}
\end{equation}
where ${\sigma}^{\mu}_{A {\dot{B}}}$
are the unit $2 \time 2$ matrix for $\mu = 0$, and the three hermitian Pauli matrices for $\mu = 1,2,3$.

Again, we want to get the same form of these four matrices in another basis. Knowing
that the lower indices  $A$ and ${\dot{B}}$ undergo the transformation with 
matrices $U^{A'}_B$ and ${\bar{U}}^{{\dot{A}}'}_{\dot{B}}$, we demand that there exist some $4 \times 4$ matrices
$\Lambda^{{\mu}'}_{\nu}$ representing the transformation of lower indices by the matrices $U$ and ${\bar{U}}$:
\begin{equation}
\Lambda^{{\mu}'}_{\nu} \, \pi^{\nu}_{A {\dot{B}}} = U^{A'}_A \, {\bar{U}}^{{\dot{B}}'}_{\dot{B}}
 \pi^{{\mu}'}_{A' {\dot{B}}'},
\label{pitransform1}
\end{equation}
This defines the vector ($4 \times 4)$ representation of the Lorentz group. The system (\ref{pitransform1}) contains four
groups of four equations each, fgollowing the choice of values for indices $\mu'$ on one side, and the indices $A$ and $B$.
We shall show explicitly only the first four equations relating the $4 \times 4$ real matrices $\Lambda^{{\mu}'}_{\nu}$ with 
the $2 \times 2$ complex matrices $U^{A'}_B$  and ${\bar U}^{{\dot{A}}'}_{\dot{B}}$, corresponding to the value $\mu' = 0'$:
$$ \Lambda^{0'}_0 + \Lambda^{0'}_3 = U^{1'}_1 \, {\bar U}^{{\dot{1}}'}_{\dot{1}} 
+ U^{2'}_1 \, {\bar U}^{{\dot{2}}'}_{\dot{1}}, \; \; \; \; \; 
\Lambda^{0'}_0 - \Lambda^{0'}_3 = U^{1'}_2 \, {\bar U}^{{\dot{1}}'}_{\dot{2}} 
+ U^{2'}_2 \, {\bar U}^{{\dot{2}}'}_{\dot{2}}, $$
\begin{equation}
\Lambda^{0'}_0 - i \Lambda^{0'}_2 = U^{1'}_1 \, {\bar U}^{{\dot{1}}'}_{\dot{2}} 
+ U^{2'}_1 \, {\bar U}^{{\dot{2}}'}_{\dot{2}}, \; \; \; \; \; 
\Lambda^{0'}_0 + i \Lambda^{0'}_2= U^{1'}_2 \, {\bar U}^{{\dot{1}}'}_{\dot{1}} 
+ U^{2'}_2 \, {\bar U}^{{\dot{2}}'}_{\dot{1}} 
\label{Lorentz4}
\end{equation}
The next three groups of four equations are similar to the above.

With the invariant ``spinorial metric" in two complex dimensions, $\varepsilon^{AB}$ and $\varepsilon^{{\dot{A}}{\dot{B}}}$ 
such that $\varepsilon^{12} = - \varepsilon^{21} = 1$
and $\varepsilon^{{\dot{1}}{\dot{2}}} = - \varepsilon^{{\dot{2}}{\dot{1}}}$, we can define
the contravariant components $\pi^{\nu \, \, A {\dot{B}}}$. It is easy to show that the
Minkowskian space-time metric, invariant under the Lorentz transformations, can be defined as
\begin{equation}
g^{\mu \nu} = \frac{1}{2} \biggl[ \pi^{\mu}_{A {\dot{B}}} \, \pi^{\nu \, \, A {\dot{B}}} \biggr] 
= diag (+,-,-,-)
\label{Mmetric}
\end{equation}
Together with the anti-commuting spinors ${\psi}^{\alpha}$ the four real coefficients defining
a Lorentz vector, $x_{\mu} \, {\pi}^{\mu}_{A {\dot{B}}}$, can generate now the supersymmetry
via standard definitions of super-derivations.

Let us then choose the matrices $S^{\alpha'}_{\beta}$ to be the usual spinor representation of
the $SL(2, {\bf C})$ group, while the matrices $U^{A'}_{B}$ will be defined as follows:
\begin{equation}
U^{1'}_{1} = j S^{1'}_1,   U^{1'}_{2} = - j S^{1'}_2, 
U^{2'}_{1} = - j  S^{2'}_1,  U^{2'}_{2} = j S^{2'}_2, 
\label{Umatrices}
\end{equation}
the determinant of $U$ being equal to $j^2$. Obviously, the same reasoning leads to the conjugate cubic representation of 
the same symmetry group $SL(2, {\bf C})$ if we require the covariance of the conjugate tensor 
$${\bar{\rho}}^{\dot{\beta}}_{{\dot{D}}{\dot{E}}{\dot{F}}} = j \, 
{\bar{\rho}}^{\dot{\beta}}_{{\dot{E}}{\dot{F}} {\dot{D}}}
= j^2 \, {\bar{\rho}}^{\dot{\beta}}_{{\dot{F}} {\dot{D}} {\dot{E}}},$$
by imposing the equation similar to (\ref{covtrans1})
\begin{equation}
{\bar{S}}^{{\dot{\alpha}}'}_{{\dot{\beta}}} \, {\bar{\rho}}^{{\dot{\beta}}}_{{\dot{A}}{\dot{B}}{\dot{C}}} = 
{\bar{\rho}}^{{\dot{\alpha}}'}_{{\dot{A}}' {\dot{B}}' {\dot{C}}'} {\bar{U}}^{{\dot{A}}'}_{{\dot{A}}} \, 
{\bar{U}}^{{\dot{B}}'}_{{\dot{B}}} \, {\bar{U}}^{{\dot{C}}'}_{{\dot{C}}} .
\label{covtrans2}
\end{equation}
The matrix $\bar{U}$ is the complex conjugate of the matrix $U$, and its determinant is $j$.
Moreover, the two-component entities obtained as images of cubic combinations of quarks,
 $\psi^{\alpha} = \rho^{\alpha}_{ABC} \theta^A \theta^B \theta^C$ 
 and ${\bar{\psi}}^{\dot{\beta}} = {\bar{\rho}}^{{\dot{\beta}}}_{{\dot{D}}{\dot{E}}{\dot{F}}}
{\bar{\theta}}^{\dot{D}} {\bar{\theta}}^{\dot{E}} {\bar{\theta}}^{\dot{F}} $
 should anti-commute, because their arguments do so, by virtue of (\ref{commutation2}):
$$ (\theta^A \theta^B \theta^C) ({\bar{\theta}}^{\dot{D}} {\bar{\theta}}^{\dot{E}} {\bar{\theta}}^{\dot{F}} )
= - ({\bar{\theta}}^{\dot{D}} {\bar{\theta}}^{\dot{E}} {\bar{\theta}}^{\dot{F}})(\theta^A \theta^B \theta^C)$$ 
We have found the way to derive the covering group of the Lorentz group acting on spinors via
the usual spinorial representation. The spinors are obtained as the homomorphic image of 
tri-linear combination of three quarks (or anti-quarks).
 The quarks transform with matrices $U$ (or ${\bar{U}}$ for the anti-quarks), but these matrices are not unitary: 
their determinants are equal to $j^2$ or $j$, respectively. So, quarks cannot
be put on the same footing as classical spinors; they transform under a $Z_3$-covering of the Lorentz group.

%In the spirit of the Kaluza-Klein theory, the electric charge of a particle is the
%eigenvalue of the fifth component of the generalized momentum operator:
%$$ {\hat{p}}_5 = -i \hbar \, \frac{\partial}{\partial x^5},$$
%where $x^5$ stays for the fifth coordinate.
%
%Let the observed electric charge of the proton be $e$
% and that of the electron $-e$. If we put now the following factors
%multiplying the generators $\theta^1$ and $\theta^2$:
%$$\Theta^1 = \theta^1 \; e^{- \frac{i q x^5}{3 \hbar}}, \; \; \;
%\Theta^2 = \theta^2 \; e^{\frac{2 i q x^5}{3 \hbar}}, \; \; \;$$
%The eigenvalues of the fifth component of the momentum operator are, respectively:
%$${\hat{p}} \Theta^1 = - i \hbar \partial_5 ( \theta^1 \; e^{ \frac{2 i q x^5}{3 \hbar}})
% = - \frac{q}{3} \Theta^1, \; \; \; \; 
%{\hat{p}} \Theta^2 = - i \hbar \partial_5 ( \theta^2 \; e^{- \frac{ i q x^5}{3 \hbar}})
%= \frac{2q}{3} \Theta^2$$
%The only non-vanishing products of our generators being $\theta^1 \theta^1 \theta^2$  and $\theta^1 \theta^2 \theta^2$, 
%for the admissible products of functions representing the ternary combinations we readily get:
%$${\hat{p}} \theta^1 \theta^1 \theta^2 = q \, \theta^1 \theta^1 \theta^2, \; \; \; 
%{\hat{p}} \theta^1 \theta^2 \theta^2 = 0, $$
%which correspond to the usual combinations of $ ({\bf uud}) $ and  ($ {\bf udd} $) quarks, representing two 
%baryons: the proton and the neutron.
%

\section{A $Z_3$ generalization of Dirac's equation}

Let us first underline the $Z_2$ symmetry of Maxwell and Dirac equations, which implies their hyperbolic
character, which makes the propagation possible. Maxwell's equations {\it in vacuo}
can be written as follows:
\begin{equation}
\frac{1}{c} \frac{\partial {\bf E}}{\partial t} = {\bf \nabla} \wedge {\bf B}, \; \; \; \; 
- \frac{1}{c} \frac{\partial {\bf B}}{\partial t} = {\bf \nabla} \wedge {\bf E}.
\label{Maxwell2}
\end{equation}
These equations can be decoupled by applying the time derivation twice, which in vacuum, where
$div {\bf E} = 0$ and $div {\bf B} =0$ leads to the d'Alembert
equation for both components separately:
$$\frac{1}{c^2} \frac{\partial^2 {\bf E}}{\partial t^2} - {\bf \nabla}^2  {\bf E} = 0, \; \; \; \; \; \; 
\frac{1}{c^2} \frac{\partial^2 {\bf B}}{\partial t^2} - {\bf \nabla}^2 {\bf B} = 0.$$
Nevertheless, neither of the components of the Maxwell tensor, be it ${\bf E}$ or ${\bf B}$, can propagate
separately alone. It is also remarkable that although each of the fields ${\bf E}$ and ${\bf B}$ satisfies
a second-order propagation equation, due to the coupled system (\ref{Maxwell2}) there exists a quadratic
combination satisfying the forst-order equation, the Poynting four-vector:
$$P^{\mu} = \left[ P^0, {\bf P} \right], \; \; \; 
P^0 = \frac{1}{2} \left( {\bf E}^2 + {\bf B}^2 \right), \; \; \; {\bf P} = {\bf E} \wedge {\bf B}, \; \; \; {\rm with} \; \; 
 \partial_{\mu} P^{\mu} = 0.$$
The Dirac equation for the electron displays a similar $Z_2$ symmetry, with two coupled equations
which can be put in the following form:
\begin{equation}
i \hbar \frac{\partial }{\partial t} \psi_{+} - mc^2 \psi_{+} = 
 i \hbar {\boldsymbol \sigma} \cdot {\boldsymbol \nabla} \psi_{-}, \; \; \; \; \; 
- i \hbar \frac{\partial }{\partial t} \psi_{-} - mc^2 \psi_{-} = 
- i \hbar {\boldsymbol \sigma} \cdot {\boldsymbol \nabla} \psi_{+},
\label{Diracpmcoupled}
\end{equation}
where $\psi_{+}$ and $\psi_{-}$ are the positive and negative energy components of the
Dirac equation; this is visible even better in the momentum representation:
\begin{equation}
\left[ E -mc^2 \right] \psi_{+} = c {\boldsymbol \sigma} \cdot {\bf p} \psi_{-}, \; \; \; \; \; 
\left[ -E - mc^2 \right] \psi_{-} = - c {\boldsymbol \sigma} \cdot {\bf p} \psi_{+}.
\end{equation}
The same effect (negative energy states) can be obtained by changing the direction
of time, and putting the minus sign in front of the time derivative, as suggested
by Feynman.

Each of the components satisfies the Klein-Gordon equation, obtained by successive
application of the two operators and diagonalization:
$$ \left[  \frac{1}{c^2} \frac{\partial^2}{\partial t^2} - 
{\bf \nabla}^2  - m^2 \right] \psi_{\pm} =0$$
As in the electromagnetic case, neither of the components of this complex entity
can propagate by itself; only all the components can.

Apparently, the two types of quarks, $u$ and $d$, cannot propagate freely,
but can form a freely propagating particle perceived as a fermion, only under an extra
condition: they must belong to three {\it different} species called {\it colors}; short of this
they will not form a propagating entity.

Therefore, quarks should be described by {\it three fields} satisfying a set of coupled linear equations,
with the $Z_3$-symmetry playing a similar role of the $Z_2$-symmetry in the case of Maxwell's and
Dirac's equations. Instead of the ``-" sign multiplying the time derivative, we should use the cubic root 
of unity  $j$ and its complex conjugate $j^2$ according to the following scheme:
\begin{equation}
\frac{\partial }{\partial t} \mid \psi > = {\hat H}_{12} \mid \phi >, \; \; \; \; \; 
j \frac{\partial }{\partial t} \mid \phi > = {\hat H}_{23} \mid \chi >, \; \; \; \; \;
%\begin{equation}
j^2 \frac{\partial }{\partial t} \mid \chi > = {\hat H}_{31} \mid \psi >. 
\label{threeeqs1}
\end{equation}
We do not specify yet the number of components in each state vector, nor the
character of the hamiltonian operators on the right-hand side; the three fields $\mid \psi >$, $\mid \phi >$ and $\mid \chi >$
should represent the three colors, none of which can propagate by itself.

The quarks being endowed with mass, we can
suppose that one of the main terms in the hamiltonians is the mass operator ${\hat{m}}$; and let us suppose that the
remaining parts are the same in all three hamiltonians.  This will lead to the following three equations:
$$\frac{\partial }{\partial t} \mid \psi >  - {\hat{m}} \mid \psi > = {\hat H} \mid \phi >, \; \; \; \; \;
j \frac{\partial }{\partial t} \mid \phi > - {\hat{m}} \mid \phi > = {\hat H} \mid \chi >, \; \; \; \; \;
j^2 \frac{\partial }{\partial t} \mid \chi > - {\hat{m}} \mid \chi > = {\hat H} \mid \psi >. $$
%\label{threeeqs2}
%\end{equation}
Supposing that the mass operator commutes with time derivation, by applying three times
the left-hand side operators, each of the components satisfies the same common {\it third order} equation:
\begin{equation}
\left[ \frac{\partial^3}{\partial t^3} - {\hat{m}}^3 \right] \mid \psi > = {\hat{H}}^3  \mid \psi >.
\label{thirdorderpsi}
\end{equation}
The anti-quarks should satisfy a similar equation with the negative sign for the Hamiltonian operator. 
The fact that there exist two types of quarks in each nucleon suggests that the state vectors $\mid \psi >$, $\mid \phi >$ and $\mid \chi >$
should have two components each. When combined together, the two postulates lead to the conclusion that we must have
three two-component functions and their three conjugates:
$$\begin{pmatrix} \psi_1 \cr \psi_2 \end{pmatrix}, \; \; 
\begin{pmatrix} {\bar{\psi}}_{\dot{1}} \cr {\bar{\psi}}_{\dot{2}} \end{pmatrix}, 
\; \; \; \; \; \; 
\begin{pmatrix} \varphi_1 \cr \varphi_2 \end{pmatrix}, \; \;
\begin{pmatrix} {\bar{\varphi}}_{\dot{1}} \cr {\bar{\varphi}}_{\dot{2}} \end{pmatrix}, \; \; \; \; \;
\begin{pmatrix} \chi_1 \cr \chi_2 \end{pmatrix}, \; \;
\begin{pmatrix} {\bar{\chi}}_{\dot{1}} \cr {\bar{\chi}}_{\dot{2}} \end{pmatrix},$$
which may represent three colors, two quark states (e.g. ``up" and ``down"), and two anti-quark states 
(with anti-colors, respectively).
%$$\begin{pmatrix} \psi_1 \cr \psi_2 \end{pmatrix}, \; \; \;  
%\begin{pmatrix} {\bar{\psi}}_{\dot{1}} \cr {\bar{\psi}}_{\dot{2}} \end{pmatrix};
%\; \; \; \; \; \; \; \;
%\begin{pmatrix} \varphi_1 \cr \varphi_2 \end{pmatrix}, \;  
%\begin{pmatrix} {\bar{\varphi}}_{\dot{1}} \cr {\bar{\varphi}}_{\dot{2}} \end{pmatrix} ;
%\; \; \; \; \; \; \; \;
%\begin{pmatrix} \chi_1 \cr \chi_2 \end{pmatrix}, \; \;
%\begin{pmatrix} {\bar{\chi}}_{\dot{1}} \cr {\bar{\chi}}_{\dot{2}} \end{pmatrix},$$
Finally, in order to be able to implement the action of the $SL(2, {\bf C})$ group via its $2 \times 2$ matrix representation
defined in the previous section, we choose the Hamiltonian ${\hat{H}}$ equal to the operator ${\bf \sigma} \cdot {\bf \nabla}$, the same as
in the usual Dirac equation. The action of the $Z_3$ symmetry is represented by factors $j$ and $j^2$, while the $Z_2$ symmetry between
particles and anti-particles is represented by the ``-" sign in front of the time derivative. 
The differential system that satisfies all these assumptions is as follows:
{\small
$$ - i \hbar \frac{\partial}{\partial t} \psi - mc^2  \psi = -  i \hbar c \; ( {\boldsymbol \sigma}\cdot{\boldsymbol \nabla}) {\bar{\varphi}}, \; \; \; \; 
i \hbar \frac{\partial}{\partial t} {\bar{\varphi}} - j mc^2  {\bar{\varphi}} = - i \hbar c \; ( {\boldsymbol \sigma}\cdot{\boldsymbol \nabla}) \chi, \; \; \; \; 
- i \hbar \frac{\partial}{\partial t} \chi - j^2 mc^2  \chi = -  i \hbar c \; ( {\boldsymbol \sigma}\cdot{\boldsymbol \nabla}) {\bar{\psi}},$$
\begin{equation}
  i \hbar \frac{\partial}{\partial t} {\bar{\psi}} - mc^2  {\bar{\psi}} = - i \hbar c \; ( {\boldsymbol \sigma}\cdot{\boldsymbol \nabla})\varphi, \; \; \; \; 
- i \hbar \frac{\partial}{\partial t} \varphi - j^2 mc^2  \varphi=  - i \hbar c \; ( {\boldsymbol \sigma}\cdot{\boldsymbol \nabla}){\bar{\chi}}, \; \; \; \; 
%\begin{equation}
 i \hbar \frac{\partial}{\partial t} {\bar{\chi}} - j mc^2 {\bar{\chi}} = - i \hbar c \; ( {\boldsymbol \sigma}\cdot{\boldsymbol \nabla}) \psi,
\label{sixeqs}
\end{equation}}
Here we made a simplifying assumption that the mass operator is just proportional to the identity matrix, and therefore commutes
with the operator ${\boldsymbol \sigma} \cdot {\boldsymbol \nabla}$.

The functions $\psi$, $\varphi$ and $\chi$ are related to their conjugates via the following 
third-order equations:
$$
 \left[ - i  \frac{\partial^3}{\partial t^3} -  \frac{ m^3 c^6}{{\hbar}^3} \right] \, \psi =  - i ({\bf \sigma} \cdot {\bf \nabla})^3   {\bar{\psi}} = 
 \left[ - i {\bf \sigma} \cdot {\bf \nabla} \right] (\Delta {\bar{\psi}}), $$ 
\begin{equation}
 \left[ i  \frac{\partial^3}{\partial t^3} - \frac{ m^3 c^6}{{\hbar}^3} \right] \, {\bar{\psi}} =   - i ({\bf \sigma} \cdot {\bf \nabla})^3  \psi = 
 \left[ - i {\bf \sigma} \cdot {\bf \nabla} \right] (\Delta \psi), 
\label{thirddiffeq}
\end{equation}
and the same, of course, for the remaining wave functions $\varphi$ and $\chi$.

The overall $Z_2 \times Z_3$ symmetry can be grasped much better if we use the matrix notation, encoding the
system of linear equations (\ref{sixeqs}) as an operator acting on a single vector composed of all the components.
Then the system (\ref{sixeqs}) can be written with the help of the following $6 \times 6$ matrices composed of
blocks of $3 \times 3$ matrices as follows:
\begin{equation}
{\Gamma}^0 = \begin{pmatrix} I & 0 \cr 0 & -I  \end{pmatrix}, \; \; \; \; 
B = \begin{pmatrix} B_1 & 0 \cr 0 & B_2  \end{pmatrix},\; \; \; \; 
P = \begin{pmatrix} 0 & Q \cr Q^T & 0  \end{pmatrix},
\label{threemat1}
\end{equation}
with $I$ the $3 \times 3$ identity matrix, and the $3 \times 3$ matrices $B_1, \; B_2$ and $Q$ defined as
follows:
$$B_1 = \begin{pmatrix} 1 & 0 & 0 \cr 0 & j & 0 \cr 0 & 0 & j^2 \end{pmatrix}, \; \; \; 
B_2 = \begin{pmatrix} 1 & 0 & 0 \cr 0 & j^2 & 0 \cr 0 & 0 & j  \end{pmatrix}, \; \; \; 
Q = \begin{pmatrix} 0 & 1 & 0 \cr 0 & 0 & 1 \cr 1 & 0 & 0 \end{pmatrix}.$$
%\label{threemat3}
%\end{equation}

The matrices $B_1$ and $Q$ generate the algebra of traceless $3 \times 3$ matrices with determinant $1$, introduced by
 Sylvester and Cayley under the name of {\it nonion  algebra}.
With this notation, our set of equations (\ref{sixeqs}) can be written in a very compact way:
\begin{equation}
- i \hbar \Gamma^0 \, \frac{\partial}{\partial t} \, \Psi = \left[ B m - i \hbar Q {\bf \sigma \cdot \nabla} \right] \, \Psi,
\label{matrixeq}
\end{equation}
Here  $\Psi$ is a column vector containing the six fields, $[\psi, \varphi, \chi, \; {\bar{\psi}}, {\bar{\varphi}}, {\bar{\chi}} ],$ 
in  this order. 

But the same set of equations can be obtained if we dispose the six fields in a 
$6 \times 6$ matrix, on which the
operators in (\ref{matrixeq}) act in a natural way:
\begin{equation}
\Psi = \begin{pmatrix} 0 & X_1 \cr X_2 & 0 \end{pmatrix} , \; \; {\rm with} \; \; \; \; 
X_1 = \begin{pmatrix} 0 & \psi & 0 \cr 0 & 0 & \phi \cr \chi & 0 & 0  \end{pmatrix},
\; \; \; \; 
X_2 = \begin{pmatrix} 0 & 0 & {\bar{\chi}} \cr {\bar{ \psi}} & 0 & 0 \cr 0 &  {\bar{\varphi}} & 0
 \end{pmatrix} 
\label{matricesQ}
\end{equation}
By consecutive application of these operators we can separate the variables and find the common equation of sixth order
that is satisfied by each of the components:
\begin{equation}
- {\hbar}^6 \frac{\partial^6}{\partial t^6} \psi - m^6 c^{12} \psi = - {\hbar}^6 {\Delta}^3 \psi.
\label{sixthorder}
\end{equation}
Identifying quantum operators of energy and momentum,
$ - i {\hbar} \frac{\partial}{\partial t} \rightarrow E, \; \; \; - i {\hbar} {\bf \nabla} \rightarrow {\bf p},$
we can write (\ref{sixthorder}) simply as follows:
\begin{equation}
E^6 - m^6 c^{12} = \mid {\bf p} \mid^6 c^6.
\label{Ep_relation}
\end{equation}
This equation can be factorized showing how it was obtained by subsequent action of 
the operators of the system, 
(\ref{sixeqs}):
{\small
$$
E^6 - m^6 c^{12} = (E^3 - m^3 c^6)(E^3 + m^3 c^6) =$$
$$
 (E - mc^2)(jE - mc^2)(j^2 E - mc^2)(E + mc^2)(jE + mc^2)(j^2 E + mc^2) = \mid {\bf p} \mid^6 c^6.
$$}
The equation (\ref{sixthorder}) can be solved by separation of variables; the time-dependent
and the space-dependent factors have the same structure:
$$A_1 \,e^{\omega\,t} + A_2 \,e^{j \,\omega\,t} + A_3 e^{j^2 \,\omega\,t},\,
\ \ \ \ B_1\,e^{{\bf k.r}} + B_2\,e^{j\,{\bf k.r}} + B_3\,e^{j^2\,{\bf k.r}}$$
with $\omega$ and ${\bf k}$ satisfying the following dispersion relation:
\begin{equation}
\frac{\omega^6}{c^6} = \frac{m^6 c^6}{{\hbar}^6} + \mid {\bf k} \mid^6,
\label{dispersion6}
\end{equation} 
where we have identified  $E = {\hbar \omega}$  and ${\bf p} = {\hbar} {\bf k}$. 
The relation (\ref{dispersion6})
%$$\frac{\omega^6}{c^6} = \frac{m^6 c^6}{{\hbar}^6} + \mid {\bf k} \mid^6,$$
is invariant under
the action of $Z_2 \times Z_3 = Z_6$ symmetry, because to any solution with given real
$\omega$ and ${\bf k}$ one can add solutions with $\omega$ replaced by  $j \omega$ or $j^2 \omega$, $j {\bf k}$ 
or $j^2 {\bf k}$, as well as $- \omega$; there is no need to introduce also $- {\bf k}$ instead of ${\bf k}$ because the vector
${\bf k}$ can take on all possible directions covering the unit sphere.
The nine complex solutions can be displayed in two $3 \times 3$ matrices as follows:
%\begin{equation}
$$\begin{pmatrix}  e^{\omega\,t - {\bf k \cdot r}} & e^{\omega\,t - j {\bf k \cdot r}}
& e^{\omega\,t - j^2 {\bf k \cdot r}} \cr 
e^{j \omega\,t-{\bf k \cdot r }} & 
e^{j \omega\,t - j {\bf k \cdot r}} &
e^{j \omega\,t - j^2 {\bf k \cdot r}} \cr
e^{j^2 \omega\,t -{\bf k \cdot r}}& e^{j^2 \omega\,t - {\bf k \cdot r}}  & 
e^{j^2 \omega\,t - j^2 {\bf k \cdot r}} 
\end{pmatrix}, \; \; \; \; \; \; \; 
\begin{pmatrix} e^{- \omega\,t - {\bf k \cdot r}} & e^{-\omega\,t - j {\bf k \cdot r}}
& e^{-\omega\,t - j^2 {\bf k \cdot r}} \cr 
e^{-j \omega\,t-{\bf k \cdot r }} & 
e^{-j \omega\,t - j {\bf k \cdot r}} &
e^{-j \omega\,t - j^2 {\bf k \cdot r}} \cr
e^{-j^2 \omega\,t -{\bf k \cdot r}}& e^{-j^2 \omega\,t - {\bf k \cdot r}}  & 
e^{-j^2 \omega\,t - j^2 {\bf k \cdot r}} 
\end{pmatrix}$$
and their nine independent  products can be represented in a basis of real functions as
{\small
$$
\begin{pmatrix}   e^{\omega\,t - {\bf k \cdot r}} &   e^{\omega\,t + \frac{{\bf k \cdot r}}{2}}
\, \cos ({\bf k} \cdot {\bf \xi}) &  e^{\omega\,t + \frac{{\bf k \cdot r}}{2}} \, \sin ({\bf k} \cdot {\bf \xi}) \cr 
 e^{- \frac{\omega\,t}{2}-{\bf k \cdot r }} \, \cos \omega \tau & 
e^{- \frac{\omega\,t}{2}+ \frac{\bf k \cdot r}{2} } \, \cos (\omega \tau - {\bf k} \cdot {\bf \xi}) &
e^{- \frac{\omega\,t}{2}+ \frac{{\bf k \cdot r}}{2}} \, \cos (\omega \tau  + {\bf k} \cdot {\bf \xi}) \cr
e^{- \frac{\omega\,t}{2} -{\bf k \cdot r}} \, \sin \omega \tau & 
e^{- \frac{\omega\,t}{2}+ \frac{ {\bf k \cdot r}}{2}} \, \sin (\omega \tau + {\bf k} \cdot {\xi}) & 
e^{- \frac{\omega\,t}{2}+\frac{{\bf k \cdot r}}{2}} \, \sin (\omega \tau - {\bf k} \cdot {\bf \xi}) 
\end{pmatrix}
$$ }
where $\tau=\frac{\sqrt{3}}{2} \,t$ and  $\xi=\frac{\sqrt{3}}{2}{\bf kr}$; similarly for the conjugate solutions
(with $- \omega$ instead of $\omega$).

The functions displayed in the matrix do not represent a wave; however, one can produce a propagating
solution by forming certain cubic combinations, e.g. 
{\small
$$e^{\omega\,t - {\bf k \cdot r}} \, e^{- \frac{\omega\,t}{2} + \frac{{\bf k \cdot r}}{2}} \, \cos (\omega \tau - {\bf k} \cdot {\bf \xi})  \,
e^{- \frac{\omega\,t}{2} + \frac{{\bf k \cdot r}}{2}} \, \sin (\omega \tau - {\bf k} \cdot {\bf \xi} ) = \frac{1}{2} \,
 \sin ( 2  \omega \tau - 2 {\bf k} \cdot {\bf \xi}). $$}

 What we need now is a multiplication scheme that would define triple products of non-propagating solutions yielding
propagating ones, like in the example given above, but under the condition that the factors belong to three distinct
subsets b(which can be later on identified as ``colors"). 

This can be achieved with the $3 \times 3$ matrices
of three types, containing the solutions displayed in the matrix, distributed in a particular way, each of the three
matrices containing the elements of one particular line of the matrix:
{\small
\begin{equation}
[A]= \begin{pmatrix}  0 & A_{12} \, e^{\omega\,t - {\bf k \cdot r}} & 0 \cr 0 & 0 & A_{23} \, 
e^{\omega\,t + \frac{{\bf k \cdot r}}{2}}
\, \cos {\bf k} \cdot {\bf \xi} \cr 
A_{31} e^{\omega\,t + \frac{{\bf k \cdot r}}{2}} \, \sin {\bf k} \cdot {\bf \xi} & 0 & 0 
\end{pmatrix}
\label{bigmatrixA1}
\end{equation} }
{\small
\begin{equation}
[B]= \begin{pmatrix}  0 & B_{12} \, e^{- \frac{\omega}{2}\,t  + \frac{{\bf k \cdot \bf r}}{2} } \, \cos ( \tau + {\bf k \cdot \xi} ) & 0 \cr 
0 & 0 & B_{23} \, e^{- \frac{\omega}{2}\,t - {\bf k \cdot r}} \, \sin \tau \cr 
B_{31} e^{\omega\,t - {\bf k \cdot r}} \, \cos \tau  & 0 & 0 
\end{pmatrix}
\label{bigmatrixB}
\end{equation} }
{\small
\begin{equation}
[C] = \begin{pmatrix} 0 & C_{12} \, e^{-\frac{\omega}{2} \,t + \frac{{\bf k \cdot r}}{2}} \, 
\cos (u)& 0 \cr 
0 & 0 & C_{23} \, e^{- \frac{\omega}{2} \,t + \frac{{\bf k \cdot r}}{2}} \, 
\sin (v) \cr 
C_{31} e^{-\frac{\omega}{2} \,t + \frac{{\bf k \cdot r}}{2}} \, \cos (u) & 0 & 0 
\end{pmatrix}
\label{bigmatrixA2}
\end{equation} }
where we  have set $u = \tau + {\bf k \cdot \xi}, \; \; \; v = \tau - {\bf k \cdot \xi}$

Now it is easy to check that in the product of the above three matrices, $ABC$ all real exponentials cancel, leaving the periodic
functions of the argument $\tau + {\bf k \cdot r}$. The trace of this triple product is equal to 
$ Tr(ABC) = [\sin \tau \, \cos ({\bf k \cdot r}) + \cos \tau \, \sin ({\bf k \cdot r}) ] \, \cos (\tau + {\bf k \cdot r}) +
\cos (\tau + {\bf k \cdot r}) \, \sin (\tau + {\bf k \cdot r}),$

representing a plane wave propagating towards $-{\bf k}$. Similar solution can be obtained with the opposite direction.
From four such solutions one can produce a propagating Dirac spinor.
This model makes free propagation of a single quark impossible, (except for a very short distances
due to the damping factor), while three quarks can form a freely propagating state.
\vskip 0.4cm
\indent
{\bf Acknowledgments}

The author expresses his deep gratitude to Michel Dubois-Violette for his enlightning
suggestions and remarks.
\vskip 0.3cm

%\section*{References}

\smallskip

\end{document}